\documentclass[fleqn,usenatbib]{mnras}

\usepackage[usenames,dvipsnames]{xcolor}
\usepackage{amsmath}
\usepackage{array}
\usepackage{xfrac}
\usepackage{multirow}

\newcommand{\rev}[1]{{\textcolor{black}{#1}}}
\usepackage[T1]{fontenc}
\usepackage{ae,aecompl}
\hypersetup{draft}

\title[MaNGA metallicity distributions]{Gas metallicity distributions in SDSS-IV MaNGA galaxies: what drives gradients and local trends?} 

\author[Boardman et al.]{
N.~Boardman$^{1}$\thanks{E-mail: nfb@st-andrews.ac.uk},
V.~Wild$^{1}$,
T.~Heckman$^{2}$,
S.~F.~Sanchez$^{3}$,
R.~Riffel$^{4,5}$,
R.~A.~Riffel$^{5,6}$,
G.~Zasowski$^{7}$\\
$^{1}$School of Physics and Astronomy, University of St Andrews, North Haugh, St Andrews KY16 9SS, UK\\
$^{2}$The William H. Miller III Department of Physics \& Astronomy, Johns Hopkins University, Baltimore, MD 21218, United States\\
$^{3}$Universidad Nacional Autónoma de México, Instituto de Astronomıa, A.P. 70-264, 04510, Mexico, D.F., Mexico\\
$^{4}$jDepartamento de Astronomia, Instituto de F\'\i sica,
Universidade Federal do Rio Grande do Sul, CP 15051, 91501-970, Porto
Alegre, RS, Brazil \\
$^{5}$Laborat\'orio Interinstitucional de e-Astronomia - LIneA, Rua
Gal. Jos\'e Cristino 77, Rio de Janeiro, RJ - 20921-400, Brazil\\
$^{6}$Departamento de F\'isica, CCNE, Universidade Federal de Santa Maria, 97105-900, Santa Maria, RS, Brazil\\
$^{7}$Department of Physics \& Astronomy, University of Utah, Salt Lake City, UT, 84112, USA\\
}

\date{Accepted X. Received X; in original form X}
\pubyear{2022}
\begin{document} 
\label{firstpage}
\pagerange{\pageref{firstpage}--\pageref{lastpage}}
\maketitle
%\nocite{*}

\begin{abstract}

The gas metallicity distributions across individual galaxies and across galaxy samples can teach us much about how galaxies evolve. Massive galaxies typically possess negative metallicity gradients, and mass and metallicity are tightly correlated on local scales over wide range of galaxy masses; however, the precise origins of such trends remain elusive. Here, we employ data from SDSS-IV MaNGA to explore how gas metallicity depends on local stellar mass density and on galactocentric radius within individual galaxies. We also consider how the strengths of these dependencies vary across the galaxy mass-size plane. We find that radius is more predictive of local metallicity than stellar mass density in extended lower mass galaxies, while we find density and radius to be almost equally predictive in higher-mass and more compact galaxies. Consistent with previous work, we find a mild connection between metallicity gradients and large-scale environment; however, this is insufficient to explain variations in gas metallicity behaviour across the mass-size plane. We argue our results to be consistent with a scenario in which extended galaxies have experienced smooth gas accretion histories, producing negative metallicity gradients over time. We further argue that more compact and more massive systems have experienced increased merging activity that disrupts this process, leading to flatter metallicity gradients and more dominant density-metallicity correlations within individual galaxies.   

\end{abstract}


\section{Introduction}\label{intro}

Gas-phase chemical abundances in galaxies are products of a range of physical processes; thus, these abundances can teach us much about how galaxies form and evolve. Modern integral-field unit (IFU) datasets -- including CALIFA \citep{sanchez2012}, SAMI \citep{croom2012}, and MaNGA \citep{bundy2015}-- provide spatial distributions of abundances across large numbers of galaxies and have led to a range of advances over the preceding decade.

Notably, the majority of massive galaxies possess negative gas metallicity gradients. \citet{sanchez2014}, for instance, report a characteristic slope of -0.1 dex per disc effective radius in their sample of CALIFA galaxies \citep[see also][]{sanchez2012b}, and \citet{sm2018} report a similar slope in their MUSE galaxy sample. Gas metallicity gradients have been reported from observations to correlate with a range of other galaxy properties including stellar mass \citep{belfiore2017, mingozzi2020, schaefer2020,franchetto2021}, size \citep{carton2018}, gas fraction \citep{franchetto2021}, bulge-to-total (B/T) ratio \citep{moran2012}, and environment  \citep{lian2019,franchetto2021}. \citet{boardman2021} report a tight trend in gradients across the mass-size plane for MaNGA galaxies, wherein more extended galaxies display more strongly negative gradients at a given stellar mass; this is similar to what was reported for stellar metallicity gradients by \citet{li2018}, and is found in \citet{boardman2021} to be significantly tighter than gradient trends with mass or size individually. Such a result is also analogous to the apparent complex connection between metallicity gradient, mass and morphology \citep{sanchez2021}, due to the tight trend between galaxy size and morphology at a given stellar mass \citep[e.g.][]{fl2013}.

A number of local $\sim$kpc-scale relations involving gas metallicity have also been reported in the literature. An example of this is the resolved mass-metallicity relation (rMZR) between gas metallicity and stellar mass surface density $\Sigma_*$ \citep[e.g.][]{ro2012,sanchez2013,bb2016}, which can be understood as reflecting a tight interplay between gas metallicities and local conditions \citep[e.g.][]{bb2018}, though an additional metallicity dependence on stellar mass is also apparent for lower-mass galaxies in particular \citep[e.g.,][]{bb2016,gao2018,hwang2019}. \citet{bb2016} argue that the rMZR alone is sufficient to reproduce metallicity gradients over stellar masses between $10^9$ and $10^{11} M_\odot$, while \citet[][hereafter B22]{boardman2022} find local relations -- involving metallicity, $\Sigma_*$, stellar mass $M_*$, stellar age and the D4000 index \citep{bruzual1983} -- to predict gradient trends across the mass-size plane qualitatively similar to what is observed.  Thus, gas metallicity gradients can potentially be understood as arising from local conditions across a galaxy, with inside-out formation naturally producing negative metallicity gradients in star-forming galaxies \citep[e.g.][]{franchetto2021}. 

Chemical evolution models can provide us with additional insight into gas metallicities. In the simple ``closed-box" model \citep{schmidt1963}, the interstellar medium (ISM) is enriched by successive stellar generations and the gas metallicity depends solely on the local gas fraction, with higher metallicities associated with lower gas fractions. More complex models also consider the effects of inflow and outflow, at which point the local escape velocity also becomes relevant \citep[e.g.][]{lilly2013,zhu2017}. Recent observational studies have generally favoured both gas fraction and escape velocity as being important in shaping the gas metallicity \citep{moran2012,carton2015,bb2018}. The rMZR provides a tighter correlation overall than do gas fraction or escape velocity, however, which could potentially be due to $\Sigma_*$ encoding information on both of these parameters \citep{bb2018}.

Typically, studies of the rMZR along with other local relations consider large samples of star-forming regions observed over many separate galaxies. However, further information can be gained by considering local trends within individual galaxies. An example of this can be seen in \citet{sm2019}, who investigate residual trends between gas metallicity and star-formation rate within individual MaNGA galaxies. In particular, the opportunity exists to determine whether the rMZR is the most fundamental local relation within individual systems.

Here, we employ SDSS MaNGA data to assess the strength of the correlation between metallicity and $\Sigma_*$ in individual galaxies, which we then compare to the strength of correlation between metallicity and galactocentric radius. Our direct use of radius is motivated by the steep gas metallicity gradients measured in extended star-forming MaNGA galaxies, along with the radial dependencies in both gas mass fractions \citep[e.g.][]{carton2015} and escape velocities within galaxies. We consider the gas metallicity as given by the oxygen abundance, $\mathrm{12+log(O/H)}$. We then investigate whether our results could be explained by variations in galaxies' environments, motivated primarily by the work of \citet{franchetto2021}.

This article is structured as follows. We present our sample and explain relevant parameters in \autoref{sample}. We present our results in \autoref{results} and discuss our findings in \autoref{discussion}, before summarising and concluding in \autoref{conclusion}. Throughout this work, we assume a Kroupa IMF \citep{kroupa2001, kroupa2003} and we adopt the following standard $\Lambda$ Cold Dark Matter cosmology: $\mathrm{H_0} = 71$ km/s/Mpc, $\mathrm{\Omega_M}$ = 0.27, $\mathrm{\Omega_\Lambda}$ = 0.73.

\section{Sample and data}\label{sample}

\begin{figure*}
\begin{center}
	\includegraphics[trim = 1cm 11cm 0cm 13cm,scale=1,clip]{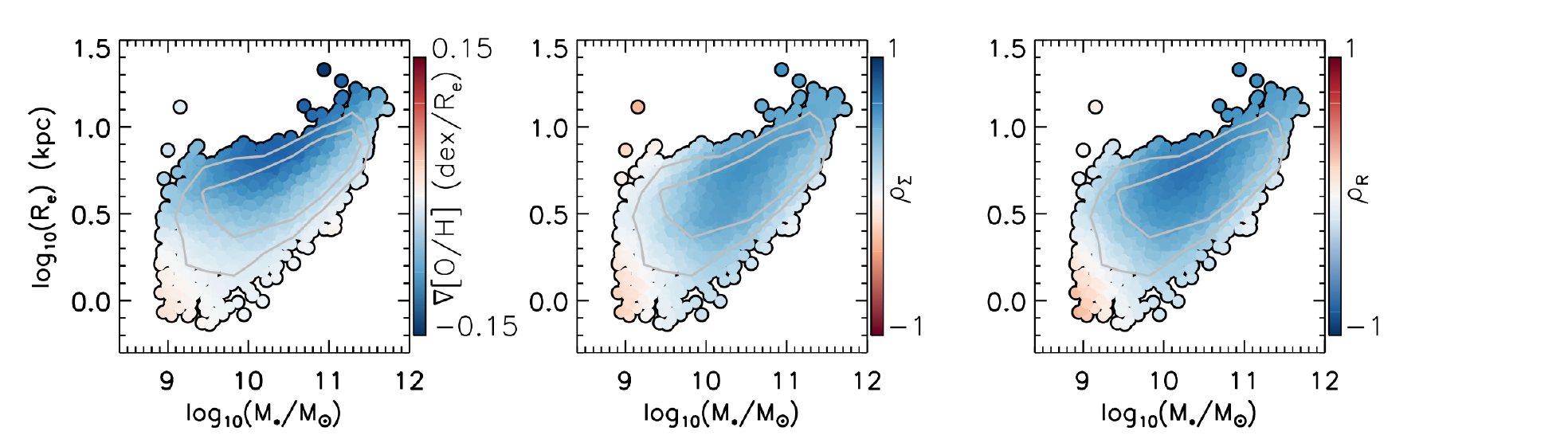}
	\caption{Effective radius, plotted against galaxy stellar mass, with data points coloured by the radial gas metallicity gradients (left) and by the Spearman correlation coefficients between $\Sigma_*$ and metallicity (middle) and between radius and metallicity (right). We apply LOESS smoothing in all panels. The middle panel uses a reversed colour scale; thus, blue regions correspond to negative gradients, positive $\Sigma_*$-metallicity correlations and negative radius-metallicity correlations. The contours enclose $\sim$50\%\ and $\sim$80\%\ of galaxies.}
	\label{rhoradgrad}
	\end{center}
\end{figure*}

We employ integral-field spectroscopic data from the SDSS-IV MaNGA survey. The MaNGA survey observed galaxies with the BOSS spectrographs \citep{smee2013} on the 2.5 meter telescope at Apache point observatory \citep{gunn2006}. The MaNGA galaxy sample covers redshifts over the approximate range of 0.01 to 0.15, and was selected to possess a roughly flat logarithmic mass distribution \citep{yan2016b,wake2017}. The MaNGA IFUs consist of bundles of 19-127 optical fibres with 2$^{\prime\prime}$ diameters \citep{law2016}. The IFUs employ hexagonal fibre configurations, and galaxies are observed with three-point dither patterns to fully sample the field of view \citep{drory2015,law2015}. The MaNGA data reduction pipeline \citep[DRP;][]{law2016,yan2016a} reduces the data, producing $0.5^{\prime\prime} \times 0.5^{\prime\prime}$ spaxel datacubes with a median PSF full-width at half-maximum (FWHM) of roughly 2.5$^{\prime\prime}$ \citep{law2016}. The MaNGA data analysis pipeline \citep[DAP;][]{belfiore2019,westfall2019} then computes various mapped quantities relating to stellar and gaseous galaxy components. MaNGA data and derived quantities are accessible through the SDSS Science Archive Server\footnote{\url{https://data.sdss.org/sas/}}, and can also be accessed through the Marvin \footnote{\url{https://www.sdss.org/dr17/manga/marvin/}} interface \citep{cherinka2019}. The reduced spectra have a resolution of $R \simeq 2000$ and cover a wavelength range of of 3600--10000\AA. The full MaNGA galaxy sample consists of more than 10000 galaxies as of SDSS DR17 \citep{sdssdr17}; however, only a subset of these will be suitable for studying gas metallicities and gas metallicity gradients. 

We selected a sample of MaNGA galaxies and spaxels in almost exactly the same manner as in B22, employing values from the DAP and from Pipe3D \citep{sanchez2016,sanchez2016b,sanchez2018}. Our one difference from B22 is that we employ the DR17 versions of Pipe3D and the DAP. We therefore refer the reader to B22 for a full description of the sample selection, though we also summarise it here. In brief, we begin by selecting MaNGA galaxies from the MPL-10 data release (which contains $\sim$9000 galaxies) with \rev{NASA-Sloan-Atlas \citep[NSA;][]{blanton2011} elliptical Petrosian axis ratios (b/a)} of at least 0.6 to avoid edge-on cases, and we cross-match with the 2X version of the GALEX-SDSS-WISE Legacy catalog \citep[GSWLC;][]{salim2016, salim2018} to obtain stellar masses. We identify star-forming spaxels with galactocentric radii between 0.5 $R_e$ and 2 $R_e$ using the BPT-NII diagnostic diagram \citep{bpt} \rev{and by requiring a minimum H$\alpha$ equivalent width of 10}, \rev{and we require a minimum S/N of 3 for the following emission features: H$\alpha$, H$\beta$, $\mathrm{[OIII]}\lambda 5007$, $\mathrm{[NII]}\lambda 6583$ and $\mathrm{[OII]}\lambda3737, 3729$}. We \rev{further} restrict to galaxies with at least 20 star-forming spaxels. We employ Pipe3d $\Sigma_*$ values along with the associated uncertainties. \rev{As in past works \citep[e.g.][]{bb2016,boardman2022}, we multiply observed $\Sigma_*$ values (along with the errors) by b/a to correct for the effects of inclination.}

\rev{We then calculate gas metallicity maps for each sample galaxy using DAP emission line fluxes with the \citet[][hereafter M13]{marino2013} calibrator. The M13 calibrator estimates gas metallicity at $\mathrm{12+log(O/H)} = 8.533 -0.214\ \mathrm{O3N2}$, where O3N2 is given as}

\begin{equation}
    \mathrm{O3N2} = \log \left(\frac{\mathrm{[OIII]}\lambda 5007}{\mathrm{H\beta}} \times \frac{\mathrm{H\alpha}}{\mathrm{[NII]}\lambda 6583}  \right)
\end{equation}

\rev{O3N2 calibrators have the advantage of being essentially unaffected by dust extinction. However, these calibrators implicitely assume a fixed N/O--O/H relation, which can bias metallicity gradient measurements in certain cases. Thus, as in B22, we present in Appendix A results from the R2 calibrator described in \citet[][]{pilyugin2016}, though we will focus on the M13 calibrator in the main paper text. As explored further in Appendix A, we obtain similar outcomes from both calibrators.}

\rev{B22 generated `model' metallicities from local relations, which they used to predict gas metallicity gradients for comparisons with observed gradients. The first of these model sets, termed `base models' in B22, constructed metallicity predictions using the three-way relation between $M_*$, $\Sigma_*$ and $\mathrm{12+log(O/H)}$ \citep[e.g.][]{gao2018}. For the second model set, termed `star formation history models' (hereafter `SFH models', B22 corrected the base models for the two-dimensional residual dependence on D4000 \citep{bruzual1983} and on light-weighted pipe3d stellar age. We construct these two model sets in the exact same manner as described in B22.} We remove spaxels in sparsely-sampled regions of the parameter space at this stage, and we remove galaxies (along with associated spaxels) that are left with fewer than 20 star-forming spaxels afterwards. Following some additional cuts as described in B22, we obtain a final sample consisting of 2127 galaxies containing \rev{860214} individual star-forming spaxels. 

The resulting sample is dominated by galaxies with significant ongoing star-formation: 2027 (95\% ) of the galaxies possess GSWLC-2X specific star formation rates (sSFRs) above $10^{-11} yr^{-1}$, while only 15 (0.7\% ) possess sSFRs below $10^{-11.5} yr^{-1}$. We calculate gas metallicity gradients for each galaxy in units of dex/$R_e$, as described in B22, where $R_e$ is the elliptical Petrosian half-light radius from the NSA catalog.

We also obtain from \citet{duckworth2019} a set of environmental parameters calculated for MaNGA galaxies. The first two of these are $\rho_{3\mathrm{Mpc}}$ and $\rho_{\mathrm{9Mpc}}$, which denote galaxy environment densities smoothed with a Gaussian kernal over local scales (3 Mpc) and larger scales (9 Mpc). In addition to these, we consider galaxies' $D_{skel}$ and $D_{node}$ parameters, which denote the distances to cosmic web nodes and filaments respectively. As in \citet{duckworth2019}, we normalise $D_{skel}$ and $D_{node}$ by the mean inter-galaxy separation at a given redshift, $\langle D_z \rangle = n(z)^{-1/3}$ where $n(z)$ is the co-moving number density; we calculate the number density as a function of redshift using the \citet{tempel2014} spectroscopic catalog of SDSS DR10 galaxies.

\begin{figure}
\begin{center}
	\includegraphics[trim = 0.5cm 10.5cm 0cm 13.5cm,scale=1]{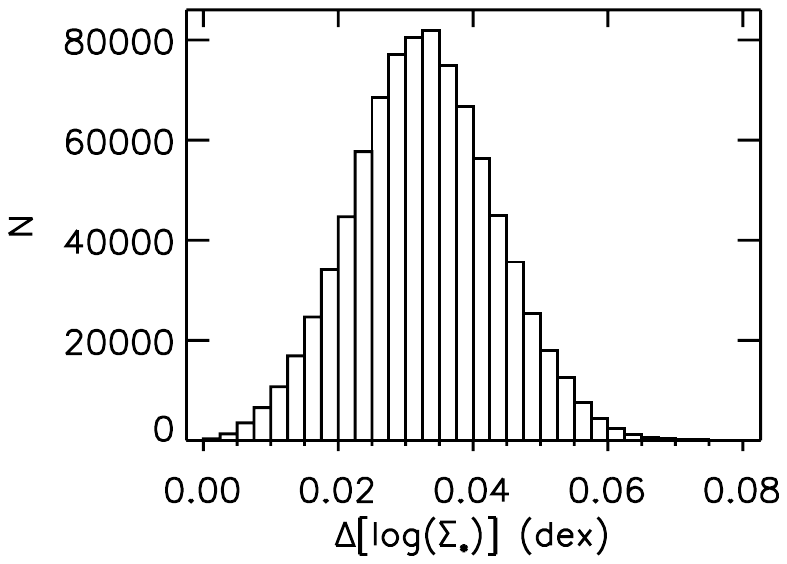}
	\caption{Histogram of errors in $\log (\Sigma_*)$.}
	\label{rhoerrs}
	\end{center}
\end{figure}

\section{Results}\label{results}

\subsection{Correlations of density and radius with gas metallicity}

We begin by considering the strength of the correlations of $\Sigma_*$ and normalised galactocentric radius $R/R_e$ with gas-phase metallicity, for spaxels in each sample galaxy separately. We parametrize this strength using the Spearman rank correlation coefficient $\rho$, computed from the IDL R\_CORRELATE procedure for each galaxy in turn; we denote the coefficient between gas metallicity and $\Sigma_*$ as $\rho_\Sigma$, while for gas metallicity and radius we use $\rho_R$. 

In \autoref{rhoradgrad}, we plot these coefficients across the mass-size plane along with the radial gas metallicity gradients ($\nabla[O/H]$) from B22. We employ locally-weighted regression smoothing \citep[LOESS;][]{cleveland1988} as implemented in IDL\footnote{available from \url{http://www-astro.physics.ox.ac.uk/~mxc/software/}} when showing these figures, to more cleanly show the overall mass-size trend. We compute the LOESS-smoothed value for each data point using the closest 20\% of data points with the \textit{rescale} keyword applied. We compute errors from the scatter in neighbouring points for the purpose of the smoothing calculation.

 It is apparent from \autoref{rhoradgrad} -- unsurprisingly so -- that the coefficients trend similarly to the gradients across the plane, with the strongest correlations seen in more massive extended galaxies for both $\Sigma_*$ and $R$. \rev{We further explore the connection between the gradients and correlation coefficients in Appendix B.} 
 
 \rev{From \autoref{rhoradgrad}, the absolute values of $\rho_R$ are typically as high or higher than $\rho_\Sigma$ However, it is not truly fair to compare $\rho_\Sigma$ and $\rho_R$ directly}, since  $\Sigma_*$ possesses non-negligible measurement errors compared to $R$. Thus, before comparing these two parameters further, we perturb the radii of spaxels to approximately match the effect of the $\Sigma_*$ errors.

\begin{figure*}
\begin{center}
	\includegraphics[trim = 0.5cm 10.5cm 0cm 13cm,scale=1,clip]{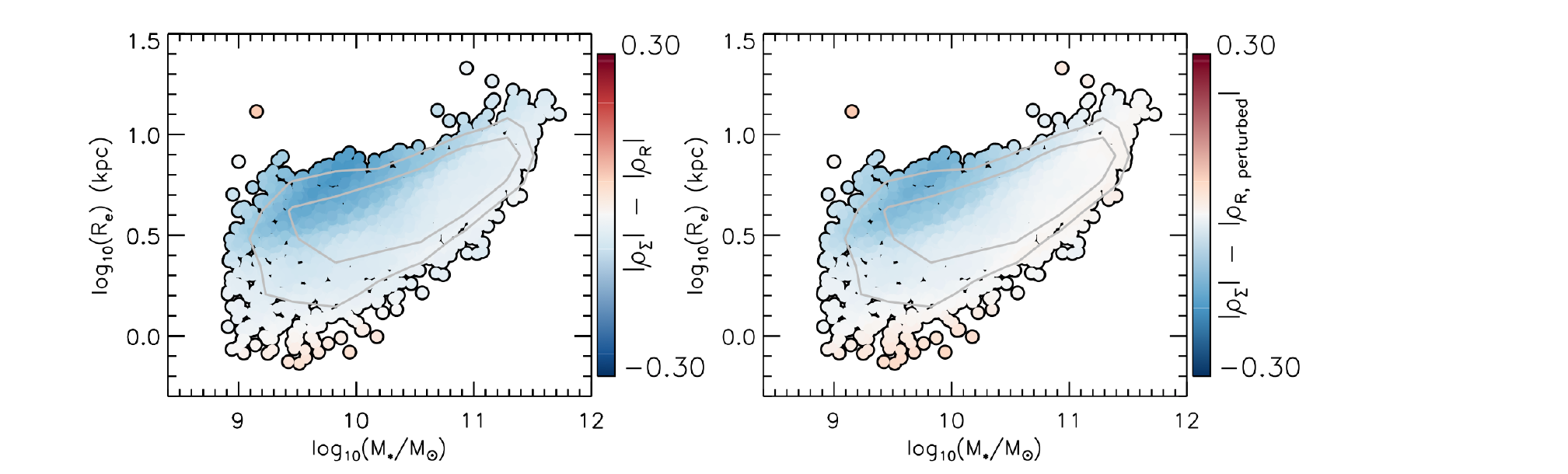}
	\caption{Difference in magnitudes of the Spearman correlation coefficient between gas metallicity and stellar density, $|\rho_{\Sigma} |$, and the coefficient between gas metallicity and galactocentric radius before ($|\rho_{R}|$; left panel) and after ($|\rho_{R, perturbed}|$; right panel) perturbing spaxel radii. Each datapoint represents one galaxy, and LOESS smoothing is applied; blue regions indicate where metallicity is more strongly correlated with radius, while red regions indicate a stronger correlation with density. The contours enclose $\sim$50\%\ and $\sim$80\%\ of galaxies.}
	\label{corrdif}
	\end{center}
\end{figure*}

\begin{figure*}
\begin{center}
	\includegraphics[trim = 1cm 11cm 0cm 13cm,scale=1,clip]{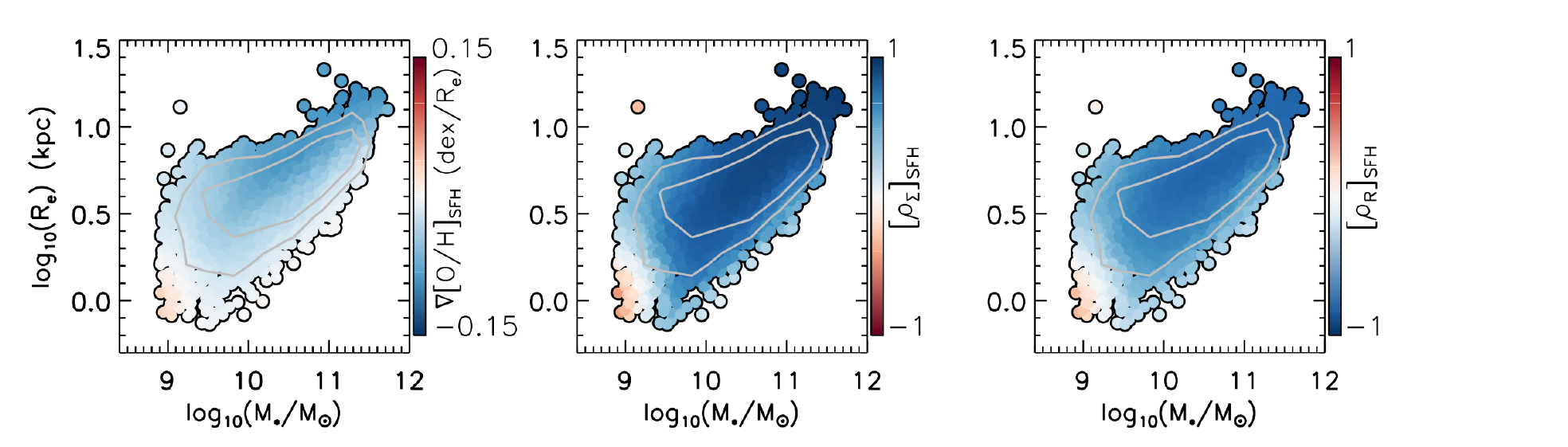}
	\caption{Effective radius, plotted against galaxy stellar mass, with data points coloured by the B22 `SFH model' metallicity gradients (left) and by the Spearman correlation coefficients between $\Sigma_*$ and model metallicity ($[\rho_{\Sigma}]_{SFH}$, middle) and between radius and model metallicity ($[\rho_{R}]_{SFH}$, right). We apply LOESS smoothing in all panels, and the middle panel uses a reversed colour scale. The contours enclose $\sim$50\%\ and $\sim$80\%\ of galaxies. We find that the SFH models do \textit{not} reproduce the trends shown in Figures 1 and 2 for observed gas metallicities: while individual parameters behave similarly across the mass-size plane, we find local metallicties to be correlated more strongly with stellar density than with radius in most cases. }
	\label{rhoradgrad_sfh}
	\end{center}
\end{figure*}

We perturb spaxels' $R/R_e$ as follows. The error in $\Sigma_*$ is typically well below 0.1 dex, as shown in \autoref{rhoerrs}; thus, we adopt 0.1 dex as a conservative estimate for the level of scatter in $\Sigma_*$ due to measurement/fitting uncertainties. We then apply a Gaussian scatter to spaxels' $\log(R/R_e)$ of $\Delta_R = 0.1 \mathrm{dex} \times \sigma_{\log(r)}/\sigma_{\log(\Sigma)}$, where $\sigma_{\log(r)}$ and $\sigma_{\log(\Sigma)}$ denote the dispersions\footnote{These dispersions and all subsequent dispersions are calculated using the ROBUST\_SIGMA IDL procedure, available at \url{https://idlastro.gsfc.nasa.gov/ftp/pro/robust/robust_sigma.pro}} in $\log(R/R_e)$ and $\log(\Sigma_*)$ over the full spaxel sample. We obtain $\sigma_{\log(r)} = 0.149\mathrm{dex}$ and $\sigma_{\log(\Sigma)} = 0.409\mathrm{dex}$, and we hence adopt $\Delta_R = 0.036\mathrm{dex}$ when perturbing spaxel radii. We calculate Spearman correlation coefficients between gas metallicity and the perturbed radii, and we denote these coefficients as $\rho_{R, perturbed}$.

\begin{figure*}
\begin{center}
	\includegraphics[trim = 0.5cm 10.5cm 0cm 13cm,scale=1,clip]{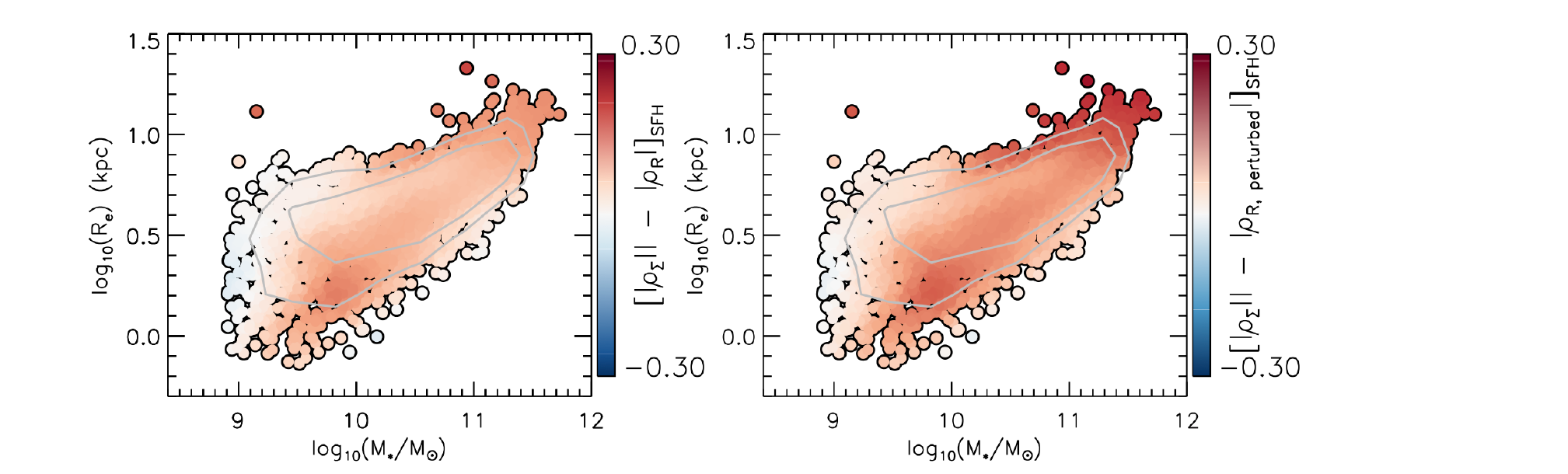}
	\caption{Difference in magnitudes of the Spearman correlation coeffcient between SFH model gas metallicity and stellar density and the coefficient between model gas metallicity and galactocentric radius, before (left panel) and after (right panel) perturbing spaxel radii. The contours enclose $\sim$50\%\ and $\sim$80\%\ of galaxies. We further demonstrate here that the B22 SFH models do not predict the behaviour of observed metallicities shown in \autoref{corrdif}.}
	\label{corrdif_sfh}
	\end{center}
\end{figure*}

In \autoref{corrdif}, we plot across the mass-size plane the difference between $|\rho_\Sigma |$ and $|\rho_{R}|$ along with the difference between $|\rho_\Sigma |$ and $|\rho_{R, perturbed}|$, with LOESS smoothing applied. We see a clear trend in the two $\rho$ values: spaxels in extended galaxies with low-to-intermediate masses have metallicities more closely correlated with R, while $\Sigma_*$ and R are almost equally predictive of the metallicity in higher-mass and more compact galaxies. We also find only mild differences in our results from using $\rho_{R}$ or $\rho_{R, perturbed}$; thus, the choice of $\rho_{R}$ or $\rho_{R, perturbed}$ does not significantly affect the final outcome.

\rev{By construction, the B22 base models cannot predict the behaviour observed in \autoref{corrdif}, due to the predicted metallicity depending entirely on $\Sigma_*$ within any given galaxy. However, the B22 SFH models depend also on D4000 and on stellar age, and so are worth considering further.} We present across the mass-size plane the resulting values of $\nabla[O/H]$, $\rho_{\Sigma}$ and $\rho_{R}$ in \autoref{rhoradgrad_sfh}, and we present the corresponding values of $|\rho_\Sigma | - |\rho_{R}|$ and $|\rho_\Sigma | -|\rho_{R, perturbed}|$ in \autoref{corrdif_sfh}. We find that the SFH models likewise do not predict the observed metallicity behaviour in these regards, with  $|\rho_\Sigma |$ greater than $|\rho_{R, perturbed}|$ over almost the whole mass-size plane once smoothing is applied. From this, we argue that radius-metallicity trends are \textit{not} simply a consequence of the local relations covered by the B22 models.

\subsection{Gas metallicity as a 2D function of density and radius.}\label{2dfunc}

\begin{figure}
\begin{center}
	\includegraphics[trim = 1cm 16.5cm 1cm 7cm,scale=1.1,clip]{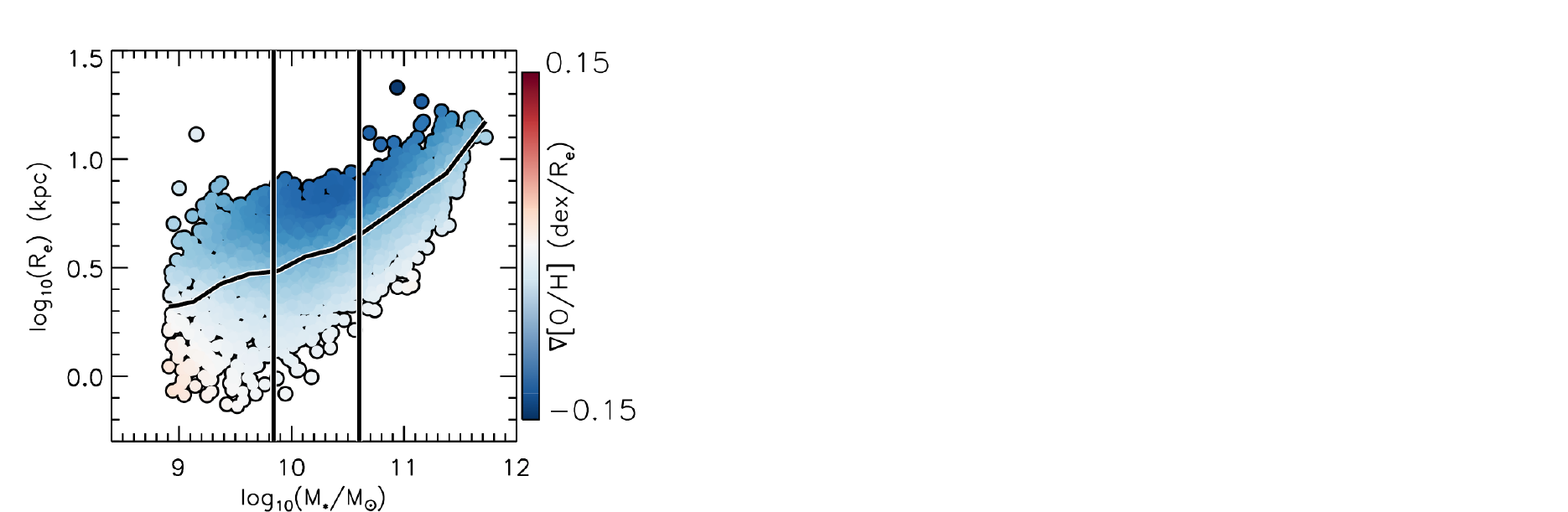}
	\caption{Same as the left window of \autoref{rhoradgrad}, but with the positions of our six mass-size bins overlaid. We split the galaxy sample into six subsamples accordingly, and consider these subsamples over the remainder of Section 4.2}
	\label{binfig}
	\end{center}
\end{figure}

To further assess how spaxel gas metallicity varies with both $\Sigma_*$ and $R$, we split our galaxy sample into subsamples based on their positions in the mass-size plane, and we then consider all spaxels within a given subsample together. Our procedure for selecting the subsamples is the same as in B22: we split the sample in stellar mass so as to encompass 1/3 of the sample apiece, and then further split the sample using the median mass-size relation calculated over a series of mass bin. We demonstrate this process in \autoref{binfig}. We deem galaxies ``extended" or ``compact" based on whether they lie above or below the median mass-size relation, and we deem galaxies ``low-mass", ``mid-mass" or ``high-mass" depending on the mass bins they fall in. 

\begin{figure*}
\begin{center}
	\includegraphics[trim = 1.5cm 11.5cm 0cm 6cm,scale=1]{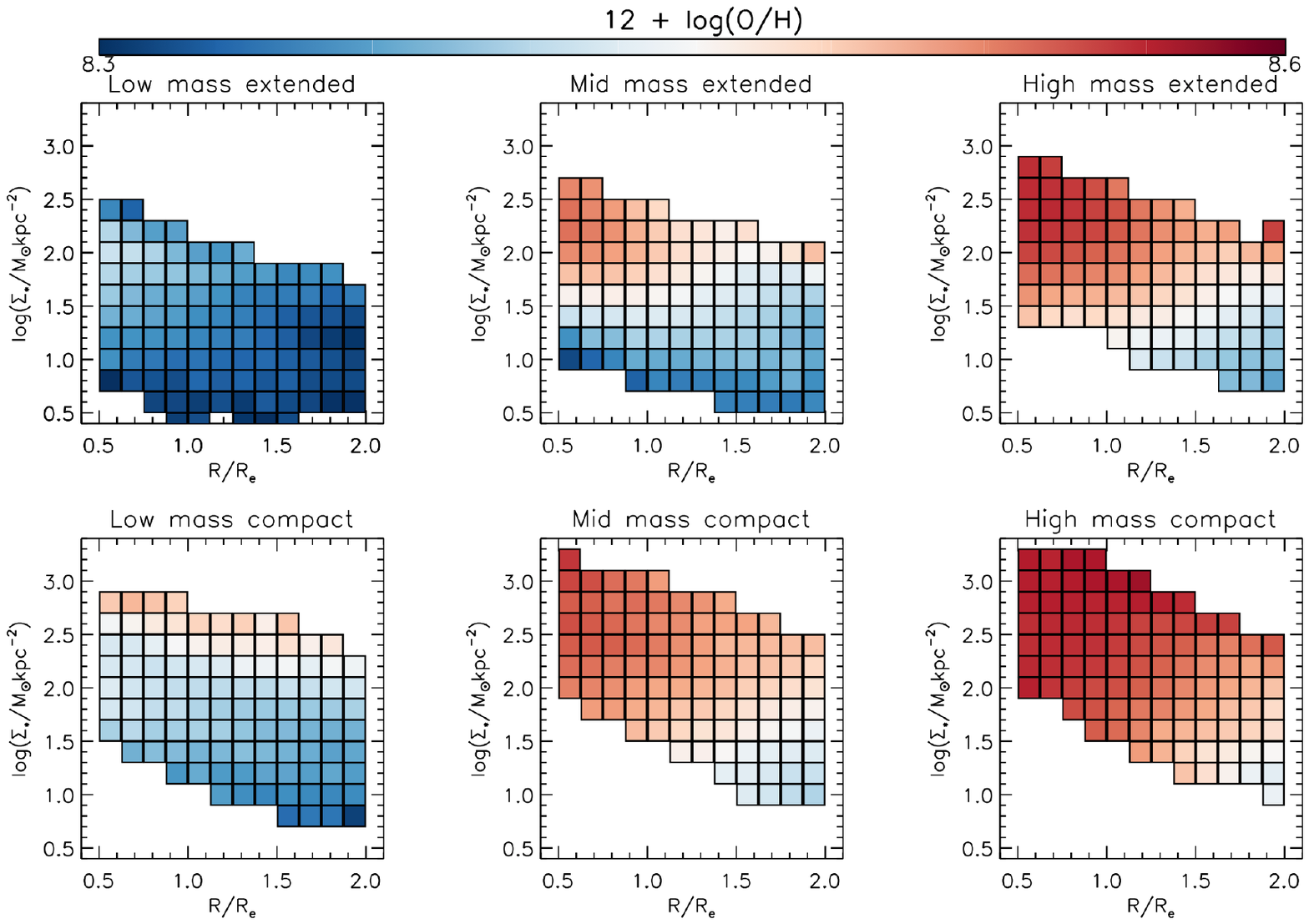}
	\caption{Mean observed spaxel gas metallicity in bins of stellar density and radius, for the six mass-size subsamples. Each bin encompasses a minimum of 50 spaxels. We find gas metallicity to vary with both $\Sigma_*$ and $R/R_e$, with the precise behaviour differing significantly across different mass-size bins.}
	\label{msbins}
	\end{center}
\end{figure*}

In \autoref{msbins} we plot the spaxel gas metallicities as a combined function of $\Sigma_*$ and $R/R_e$ for each of the six mass-size subsamples. Here, we calculate the mean metallicity in bins of $\Sigma_*$ and $R/R_e$, with all shown bins containing at least 50 spaxels. This is conceptually similar to Figure 10 of \citet{neumann2021}, which presents Firefly stellar metallicities in terms of $\Sigma_*$ and $R/R_e$ for galaxy subsamples split simultaneously by mass and morphology. In general, simultaneous splits by mass and morphology (and/or by mass and size) are a common way to explore galaxy samples in the literature \citep[e.g.][]{gb2017,sanchez2020,sanchez2021}, and allow for additional insight compared to splitting by any one property alone. We see in \autoref{msbins} that spaxel gas metallicities trend with both $\Sigma_*$ and radius, and that the shape of these trends is not consistent between different mass-size bins. Metallicity trends mainly with density in low-mass compact galaxies, for instance, while at higher masses the metallicity both rises with density and falls with radius.

\rev{Our findings here possess some differences to what \citet{neumann2021} report for stellar metallicities, In general, \citet{neumann2021} find stellar metallicity to trend chiefly with $\Sigma_*$ for most mass-morphology bins. For low-to-intermediate mass galaxies and for massive spirals, \citet{neumann2021} then find stellar. metallicity to rise with radius at fixed $\Sigma_*$. \citet{neumann2021} argue this to require a metallicity driver in addition to $\Sigma_*$, with the driver serving either to raise metallicity in galaxies' outer parts or else to dilute it in galaxies' inner regions. By contrast, we find that gas metallicity typically falls with radius at fixed $\Sigma_*$, though we also find metallicity to rise with $\Sigma_*$ at fixed radius; the only exception here is our low-mass compact galaxy subsample, which behaves similarly to what \citet{neumann2021} report for stellar metallicities.} 

\subsection{Connection with environment}

\begin{figure*}
\begin{center}
	\includegraphics[trim = 0cm 10.5cm 0.5cm 6.5cm,scale=1,clip]{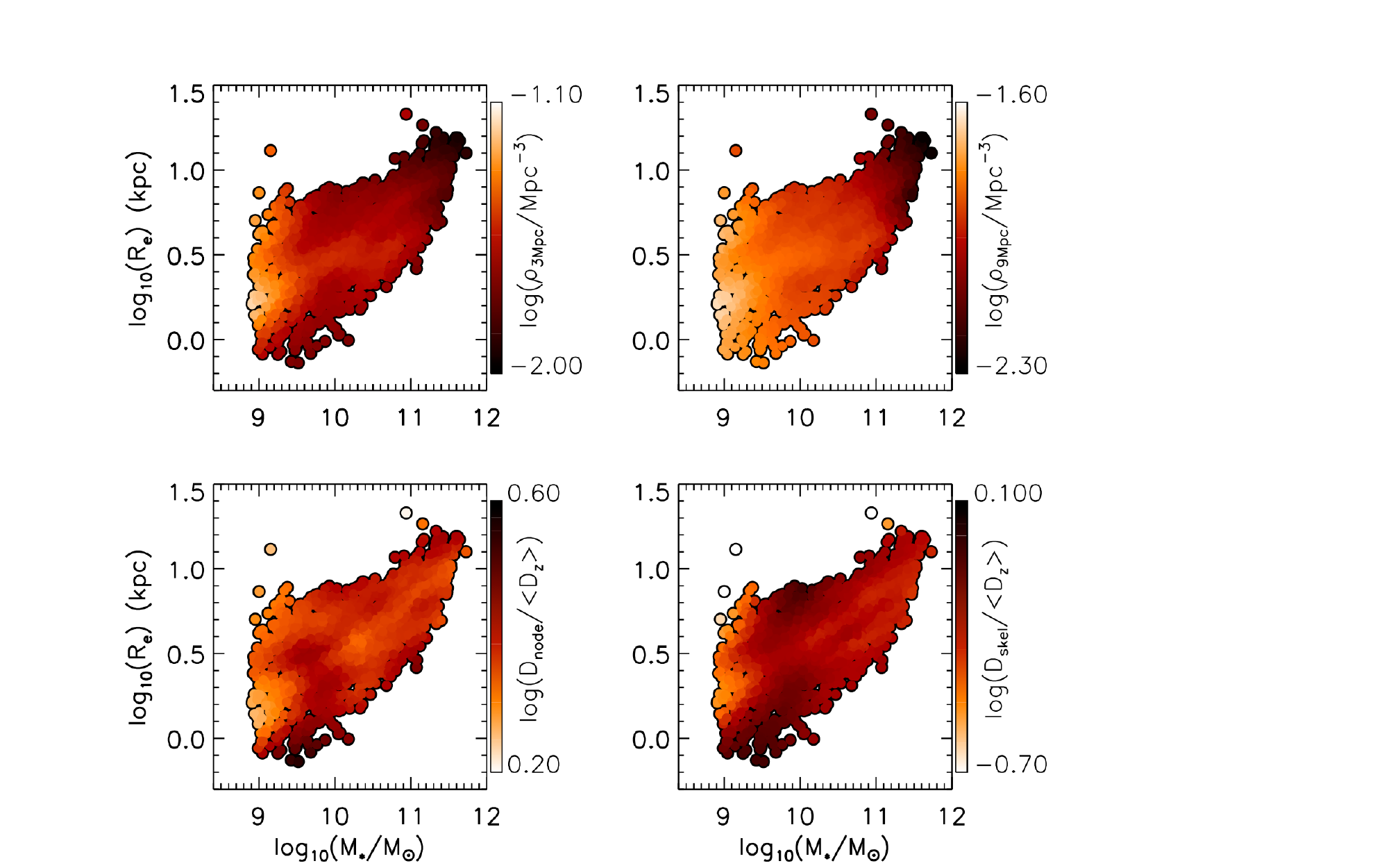}
	\caption{Environment measures plotted across the mass-size plane with LEOSS smoothing applied. Data points are coloured by local density calculated by a 3 Mpc (top left) or 9 Mpc (top right) Gaussian kernal, by the nearest node distance (bottom left) and by the nearest fillament distance (bottom right). We invert the color scale on the bottom two panels, such that darker colours are associated with less dense environments.}
	\label{msenvs}
	\end{center}
\end{figure*}

From past work, the connection between gas metallicity gradients and galaxy environment appears to be small but non-neglible. Few differences are found in the gradients of satellite and central galaxies. \citet{lian2019} only find a significant effect for satellite galaxies with $M_* < 10^{9.7} M_\odot$, for instance, for which denser environments are associated with flatter gradients. \citet{schaefer2019} likewise only detect mild gradient differences between galaxies with satellite/central classifications, with those differences depending on the chosen gas metallicity calibrator. On the other hand, cluster environments have been found to be associated with flatter gradients than are field environment at a given mass, in observations \citep{franchetto2021} as well as in simulations \citep{ll2022}. Denser environments have also been reported to be associated with smaller disc galaxy sizes, particularly at lower stellar masses \citep[e.g.][]{cebrian2014,kuchner2017}. Thus, it is worthwhile to consider whether our gas metallicity results can be explained -- at least in part -- by variations in galaxies' environments.

We now cross-match our sample with the \citet{duckworth2019} catalog, in order to assess the importance of environment for understanding our results reported up to now. We obtain environment measures for 1726 galaxies (81.1\% of our sample), and we analyse only these galaxies over the remainder of this subsection. This reduction in sample size is due to \citet{duckworth2019} employing an earlier MaNGA data-release (MPL-6) that contains fewer galaxies than MPL-10.

In \autoref{msenvs}, we present the four environment quantities described in Section 2 across the mass-size plane, with LOESS smoothing applied. We see that all four quantities primarily trend with mass, such that higher mass star-forming galaxies are associated with lower environmental densities and greater distances from cosmic web features; this is different from the trends involving metallicity gradients, in which size is found to be a significant factor. As such, we argue that environment variations are \textit{not} sufficient to explain the variation in metallicity gradients across the mass-size plane. The finding of a mass-environment connection, we note, is likely a consequence of our selected sample: the sample consists only of star-forming galaxies by design, with the most massive star-forming galaxies then being more likely to avoid quenching by residing in less dense environments.

\begin{figure*}
\begin{center}
	\includegraphics[trim = 1.5cm 10.5cm 0cm 3cm,scale=0.9,clip]{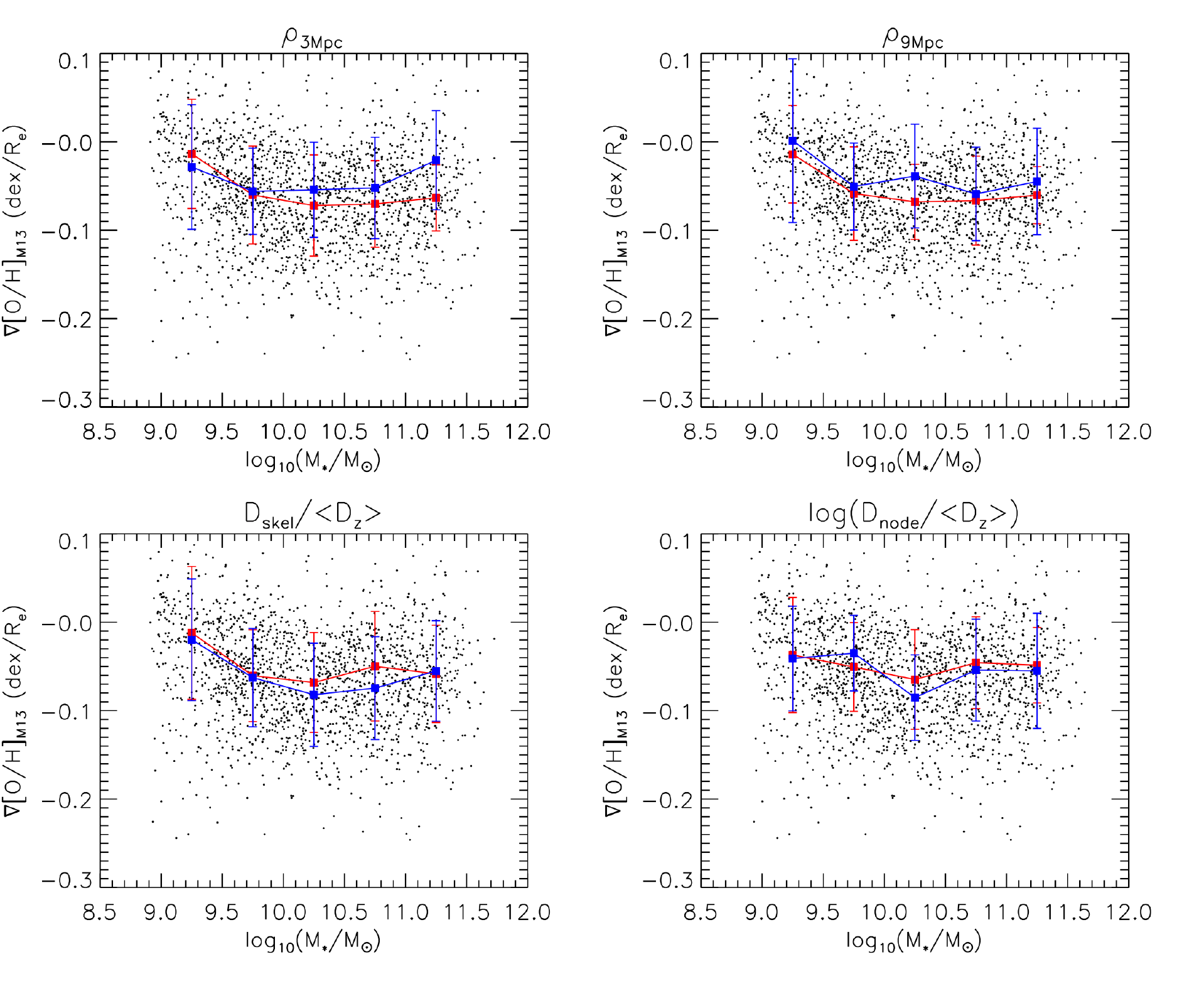}
	\caption{B22 gas metallicity gradient as a function of stellar mass. The blue (red) squares show the median gradients for galaxies in the top (bottom) 10 percentiles in a given environment measure, as indicated by the panel titles: local density with a 3 Mpc kernel (top left), local density with a 9 Mpc kernel, nearest filament distance (bottom left) and nearest node distance (bottom right).}
	\label{gradenvs}
	\end{center}
\end{figure*}

We probe for a connection between mass and environment in the following manner. We calculate the 10th and 90th percentiles of environment measures in bins of stellar mass; for each bin, we compute the median and dispersion of the gas metallicity gradient for all galaxies below the 10th percentile and above the 90th percentile separately. This results in eight sets of median and dispersion values for a given mass bin. We then plot the resulting values as a function of mass in \autoref{gradenvs}, along with showing the full environment sample. This is conceptually similar to \citet{franchetto2021}, who consider field and cluster galaxies' gas metallicity gradients as a function of stellar mass.

It is evident from \autoref{gradenvs} that galaxies in dense and sparse environments have gradients that depend differently on mass, with sparser environments -- that is, larger values of $\rho_{3Mpc}$ and $\rho_{9Mpc}$ and/or smaller values of $D_{skel}/\langle D_z \rangle$ and $D_{skel}/\langle D_z \rangle$ -- associated with a slightly stronger mass-gradient dependencies; this suggests that environment is a factor in gas metallicity gradients' scatter at a given $M_*$, regardless of how environment is quantified.

To summarise, we detect a mild connection between the metallicity properties of star-forming galaxies and those galaxies' large-scale environments. We find dense galaxy environments to be associated with slightly flatter metallicity gradients overall, consistent with the findings of \citet{franchetto2021}; thus, we find that environment is indeed relevant to the evolution of gas metallicity gradients. However, we find environment measures in our sample to trend primarily with mass, as opposed to trending with both mass and size. Thus, environment variations are \textit{not} sufficient to explain the behaviour of galaxies' gas metallicity profiles across the mass-size plane.

\section{Discussion}\label{discussion}

In the previous section, we employed MaNGA data to assess how metallicity connects with $\Sigma_*$ and galactocentric radius on local ($\sim$1 kpc) scales. Using the Spearman correlation coefficient, we found gas metallicity to be correlated more strongly with radius than with $\Sigma_*$ in lower-mass extended galaxies, \rev{while we found gas metallicity correlate roughly as strongly with $\Sigma_*$ and with radius in more massive compact galaxies}. We also considered spaxel metallicities as a combined function of radius and $\Sigma_*$, finding complex two-dimensional trends that vary with both galaxy mass and galaxy size\rev{; these trends differ somewhat from what has previously been reported for stellar metallicities \citep{neumann2021}}. Finally, we found that large-scale environment variations are \textit{not} sufficient to explain our main findings, though we do find mild differences in gas metallicity gradients for low-density and high-density galaxy environments.

Our results for lower-mass extended galaxies imply that position within the galaxy, and \textit{not} $\Sigma_*$, is the primary decider of gas metallicity within these systems. This suggests that the rMZR computed across large spaxel samples is \textit{not} sufficient to fully capture the metallicity behaviour in these particular galaxies. More complex local relations, such as those explored in B22, are likewise unable to predict this particular finding. The apparent importance of radius can potentially be explained as a result of the tight radial dependencies both on gas fraction \citep[which increases with radius; e.g.][]{carton2015} and on escape velocity \citep[which decreases with radius; e.g.][]{bb2018}, both of which influence gas metallicity evolution in chemical evolution models \citep[e.g.][]{lilly2013,zhu2017} while possessing far larger uncertainties than the radius itself. In this picture, the outer parts of galaxies evolve slower (and hence take longer to use their gas supply) and receive more inflowing metal-poor gas while also being more susceptible to metal outflows, producing negative metallicity gradient as a consequence of both points.

The above picture does not explicitly consider radial flows. Recent simulations support a view in which gas primarily accretes onto the plane of a galaxy's disc and then gradually migrates towards a galaxies' centre \citep[e.g.][]{trapp2022}. Such a picture forms the basis of the `modified accretion disc' model presented in \citet{wang2022}, which is argued in that paper to be sufficient to reproduce the observed exponential structure of star-forming discs. \citet{wang2022a} subsequently found this model to be capable of reproducing gas metallicity gradients: radially-inflowing gas continuously enriches as it travels closer to a galaxies' centre, naturally giving rise to negative metallicity gradients. However, observational evidence of significant inward gas flows remains scarce: though \citet{schmidt2016} reports significant inflow activity from direct HI observations, most such studies report results consistent with little activity \citep{wong2004,trachternach2008,teodoro2021}.

\citet{hwang2019} employed the $M_*$-$\Sigma_*$-O/H relation to identify and study regions with anomalously low gas metallicities in MaNGA galaxies; amongst other results, they found such regions to be preferentially located in galaxies' outer parts. \citet{luo2021} subsequently studied the N/O abundance ratios of star-forming gas and found anomalously-low metallicities to be associated with elevated N/O ratios at a given metallicity, with the greatest elevations seen at the largest radii; from simple models, they argue this to be evidence for metal-poor gas inflows. Such results are consistent with the \citet{wang2022a} scenario, in which metal-poor gas migrates inwards from the outer parts of a disc. More generally, these results highlight the likely importance of metal-poor inflows in understanding gas metallicity gradients, and they support a view in which inflows preferentially occur at larger radii. Such a view is entirely consistent with our findings for extended galaxies.

By contrast, more massive and more compact galaxies possess flatter gas metallicity gradients and hence display weaker metallicity-radius correlations, suggesting that other factors are relevant in shaping the metallicity distributions of these systems. One possibility is that these galaxies possess increased escape velocities due to their compactness, resulting in reduced metal outflows and a higher relative dependence on $\Sigma_*$ (which, in this picture, serves as a proxy for the gas content). However, it can be seen from \autoref{rhoradgrad} that $\rho_\Sigma$ actually weakens somewhat within the most compact massive galaxies, making it worthwhile to explore additional explanations. 

\begin{figure}
\begin{center}
	\includegraphics[trim = 1cm 16.5cm 1cm 7cm,scale=1.1,clip]{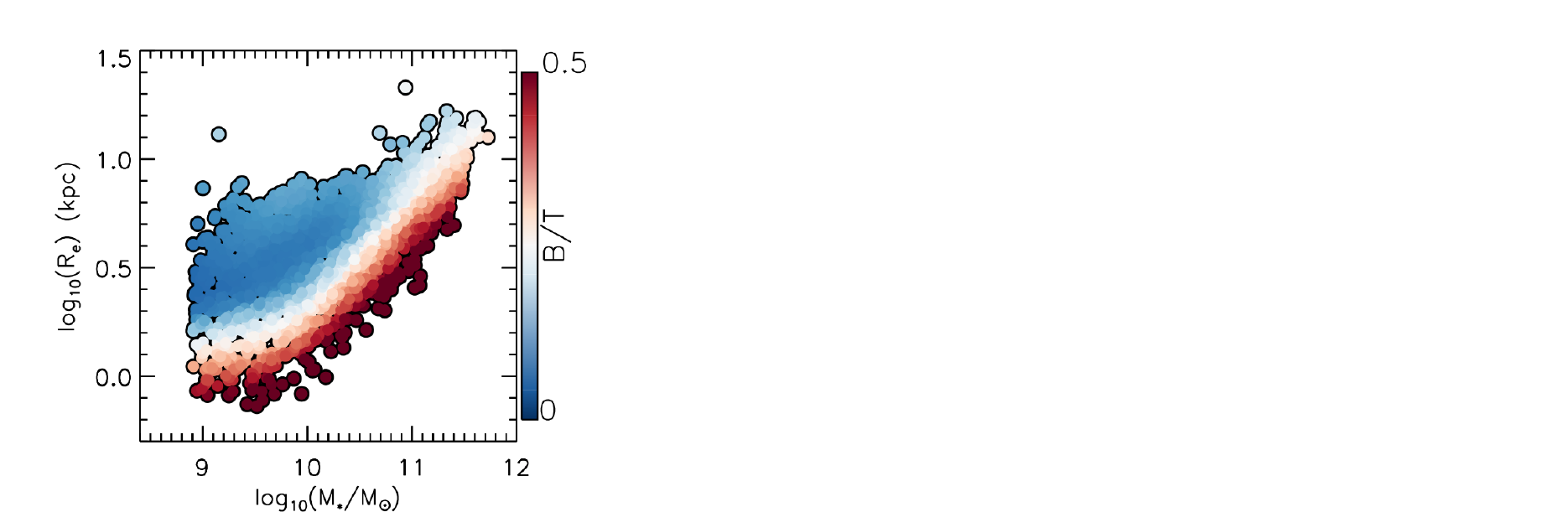}
	\caption{Galaxy bulge-to-total ratios from \citet{simard2011} plotted across the mass-size plane with LOESS smoothing applied. The color scale is such that blue regions indicate dominant disc components, with red regions indicating increasingly prominent bulges.}
	\label{btrms}
	\end{center}
\end{figure}

At a given stellar mass, compact galaxies possess earlier-type morphology than their more extended counterparts \citep[e.g.][]{fl2013,boardman2021}. To explore this point further, we cross-match our sample with the catalog of \citet{simard2011} to obtain light-weighted r-band B/T values for our galaxies, employing the fits in which the bulge Sersic index was fixed to 4. We obtain values for 2013 galaxies (94.6 \% of our full sample), and we plot these with LOESS smoothing applied in \autoref{btrms}. As expected, we find very low average B/T values amongst lower mass extended galaxies, with B/T then rising for more massive and more compact galaxies. Thus, low B/T ratios appear to be associated with strong metallicity-gradient correlations and (particularly for higher-mass galaxies) with steeper gas metallicity gradients in general. In turn, higher B/T ratios are then associated with flatter gradients and a weaker metallicity-radius dependence.

\citet{fu2013} have previously reported a significant connection between gas metallicity gradients and B/T mass ratios: in both semi-analytic models and data \citep{moran2012}, smaller B/T values are associated with steeper gas metallicity gradients. \citet{fu2013} ascribe this to differences in galaxies' merger histories: model galaxies with steep metallicity gradients are those which have continuously accreted gas in an undisturbed way, while model galaxies with shallow/flat gradients have experienced greater merging activity. In the \citet{fu2013} models, mergers trigger central starbursts and grow galaxies' bulges while also flattening metallicity gradients, with gradients then gradually steepening once smooth gas accretion resumes. \citet{fu2013} also consider the effects of inward radial flows, but find them to have little impact on gradients. It should be noted that \citet{yates2021} find no such B/T-gradient dependence in their own semi-analytic models, which employ a later version of the L-Galaxies code used by \citet{fu2013}; however, \citet{yates2021} select their sample to include only disk-dominated (B/T < 0.3) galaxies, which is a possible factor in their different findings.

Concerning the impact of mergers on gas metallicity profiles, various hydrodynamical simulations have likewise found merging/interaction activity to disrupt and flatten metallicity gradients \citep[e.g.][]{rupke2010,torrey2012}. Such a picture also has some observational support, with \citet{kewley2010} reporting flatter gradients in close galaxy pairs. A connection between mergers and bulge-growth has also been reported in simulations \citep[e.g.][]{hopkins2010,sauvaget2018}. Mergers have been associated in simulations with spin reorientation \citep{welker2014}, with galaxies transitioning from parallel to perpendicular with respect to their nearest fillimant; a similar observational effect has been reported by \citet{barsanti2022} between spin alignment and bulge mass \citep[see also][concerning the connection between alignment and morphology]{kraljic2021}. Thus, we find it plausible that different merging histories can explain differences in metallicity gradient behaviour across the mass-size plane, particularly amongst more massive galaxies. 

As has been previously noted for MaNGA galaxies, we see a flattening in gas metallicity gradients at low stellar masses. Such galaxies are metal-poorer overall than their higher-mass counterparts \citep[e.g.,][]{lequeux1979,tremonti2004}, and consequently can be expected to have evolved at a slower rate \citep{asari2007,cf2021,cf2022}. Increased metal outflows likely play a role in the flattening of low mass galaxies' gradients \citep[e.g.][]{belfiore2019a}, along with wind recycling \citep[e.g.][and references therein]{belfiore2017}. Steller metallicity distributions for low-mass spirals also appear consistent with little metal-poor gas accretion, which is not the case for higher-mass spirals \citep{greener2021}. In addition, low-mass MaNGA galaxies have been found to possess flatter gas density profiles than do higher-mass objects \citep{bb2022}, possibly due to shallower gravitational potentials in low-mass objects. Thus, we argue that a reduced prominence of metal-poor inflows \textit{and} an increased prominence of metal outflows, along with shallower gravitational potentials more generally, are what lead to flatter gas metallicity gradients in low-mass galaxies.

A potential concern in this study is the effects of beam smearing from the MaNGA PSF, particularly for galaxies observed with 19-bundle and 37-bundle fibre-IFUs. Such galaxies have the smallest spatial extents on the sky, and \citet{belfiore2017} have shown this to be associated with flatter calculated gas metallicity gradients in MaNGA galaxies. With this in mind, \citet{boardman2021} considered how MaNGA gas abundance varied with both angular (in arcseconds) and physical (in kpc) galaxy size and found significant trends between gas metallicity and \textit{physical} size at most given angular sizes; \citet{boardman2021} thus argued that gradient trends across the mass-size plane are not a result of MaNGA PSF effects, and we likewise argue that PSF effects are not the cause of our results here. The importance of PSF effects could be minimised by cutting explicitly on angular galaxy size; however, this would bias our sample towards intrinsically larger galaxies \citep{wake2017} and would reduce the coverage of our sample across the mass-size plane, so we do not consider such an approach to be optimal for our purposes.  

The overall interpretation we advocate is as follows. At any given mass, the most extended galaxies are those with the smoothest gas accretion histories. These galaxies form in an inside-out manner, resulting in steep metallicity gradients and large stellar discs. More compact galaxies have experienced more significant merging activity, by contrast, and so possess more significant bulges along with flatter metallicity gradients due to disruption of previously-established metallicity profiles. Finally, the least massive galaxies are more susceptable to significant metal outflows while also experiencing less metal-poor inflow (as evidenced by their abundant low-metallicity stars), resulting in flatter gradients overall. 

\section{Summary and conclusion}\label{conclusion}

In this paper, we have employed MaNGA data to study local gaseous and stellar metallicity as functions of stellar mass surface density and of galactocentric radius. We considered star-forming spaxels across each individual sample galaxy in turn, while also considering spaxels across galaxies grouped according to their position within the mass--size plane. 

Amongst lower mass extended galaxies, we found spaxels' gas metallicity to correlate more strongly with galactocentric radius than with density. The local relations considered in B22 -- which do not include radius -- are not able to replicate this finding. Thus, we argue that radius is an important parameter in its own right for understanding gas metallicities on local scales: metallicities are strongly related to the radius, or else are related to one or more parameters strongly correlated with the radius. We argued our findings to be consistent with a view in which extended galaxies experienced comparatively smooth gas accretion histories, with metal-poor inflows and metal outflows both preferentially affecting the galaxies' outer parts; this, along with the inside-out evolution of galaxies, naturally gives rise to negative metallicity gradients. Our results here could also be explained by the presence of significant inward radial flows, as formalised for instance by the `modified accretion disc' framework \citep[][]{wang2022,wang2022a}, but observational support for this view remains limited at present.

More compact higher-mass galaxies, meanwhile, possess a reduced correlation between gas metallicity and radius -- something that is to be expected given their relatively flattened gradients. In this case, stellar density provides a stronger correlation than radius once the effect of uncertainties are considered. This difference is associated with earlier-type morphologies -- that is, more prominent bulges -- and we argue this to reflect the impact of different merger histories for galaxies of different morphologies. 

We also investigated the impact of environment, including position within the Cosmic Web, finding only a limited connection between gas metallicity gradients and local environment. We also detected no significant connections between environment and galaxy size. We therefore argued that environment variations are \textit{not} sufficient to explain the findings described above.

Various potential extensions to this work remain. A machine learning approach \citep[e.g.][]{bluck2019,bluck2020} that includes spaxel radii along with other local parameters would allow for a more thorough investigation of local relations. Chemical evolution modelling for different classes of galaxy across the mass-size plane could also prove fruitful. 

\section*{Acknowledgements}

For the purpose of open access, the author has applied a Creative Commons Attribution (CC BY) licence to any Author Accepted Manuscript version arising. The support and resources from the Center for High Performance Computing at the University of Utah are gratefully acknowledged. NFB and VW acknowledge STFC grant ST/V000861/1.

 Funding for the Sloan Digital Sky Survey IV has been provided by the Alfred P. Sloan Foundation, the U.S. Department of Energy Office of Science, and the Participating Institutions. SDSS-IV acknowledges support and resources from the Center for High-Performance Computing at the University of Utah. The SDSS web site is \url{www.sdss.org}. J.B-B thanks IA-100420 (DGAPA-PAPIIT, UNAM) and CONACYT grant CF19-39578 support. RR thanks Conselho Nacional de Desenvolvimento Cient\'{i}fico e Tecnol\'ogico  ( CNPq, Proj. 311223/2020-6,  304927/2017-1 and 400352/2016-8), Funda\c{c}\~ao de amparo 'a pesquisa do Rio Grande do Sul (FAPERGS, Proj. 16/2551-0000251-7 and 19/1750-2), Coordena\c{c}\~ao de Aperfei\c{c}oamento de Pessoal de N\'{i}vel Superior (CAPES, Proj. 0001). RAR acknowledges financial support from Conselho Nacional de Desenvolvimento Cient\'ifico e Tecnol\'ogico (302280/2019-7).

SDSS-IV is managed by the Astrophysical Research Consortium for the Participating Institutions of the SDSS Collaboration including the Brazilian Participation Group, the Carnegie Institution for Science, Carnegie Mellon University, the Chilean Participation Group, the French Participation Group, Harvard-Smithsonian Center for Astrophysics, Instituto de Astrof\'isica de Canarias, The Johns Hopkins University, Kavli Institute for the Physics and Mathematics of the Universe (IPMU) / University of Tokyo, Lawrence Berkeley National Laboratory, Leibniz Institut f\"ur Astrophysik Potsdam (AIP),  Max-Planck-Institut f\"ur Astronomie (MPIA Heidelberg), Max-Planck-Institut f\"ur Astrophysik (MPA Garching), Max-Planck-Institut f\"ur Extraterrestrische Physik (MPE), National Astronomical Observatories of China, New Mexico State University, New York University, University of Notre Dame, Observat\'ario Nacional / MCTI, The Ohio State University, Pennsylvania State University, Shanghai Astronomical Observatory, United Kingdom Participation Group, Universidad Nacional Aut\'onoma de M\'exico, University of Arizona, University of Colorado Boulder, University of Oxford, University of Portsmouth, University of Utah, University of Virginia, University of Washington, University of Wisconsin, Vanderbilt University, and Yale University.

\section*{Data Availability}

All non-MaNGA data used here are publically available, as are all MaNGA data as of SDSS DR17.

\bibliographystyle{mnras}
\bibliography{bibliography}

\begin{appendix}

\section{Results with R2 calibrator}\label{appendix_p16}

\rev{Thus far, we have focussed on gas metallicity results from the O3N2 calibrator of \citet{marino2013}. This calibrator has the advantage of simplicity, due to it being essentially unaffected by dust attenuation. However, different calibrators can lead to notably different outcomes in terms of gas metallicities and gas metallicity gradients \citep[e.g.][]{kewley2008,schaefer2020}, and the O3N2 calibrator implicitely assumes a fixed N/O-O/H relation which, in practice, will not be true for all spaxels \citep[e.g.][]{luo2021}. As such, we briefly present results from the R2 calibrator of \citet[hereafter P16][]{pilyugin2016}, as was done in Appendix B of B22.}

\begin{figure*}
\begin{center}
	\includegraphics[trim = 1cm 11cm 0cm 13cm,scale=1,clip]{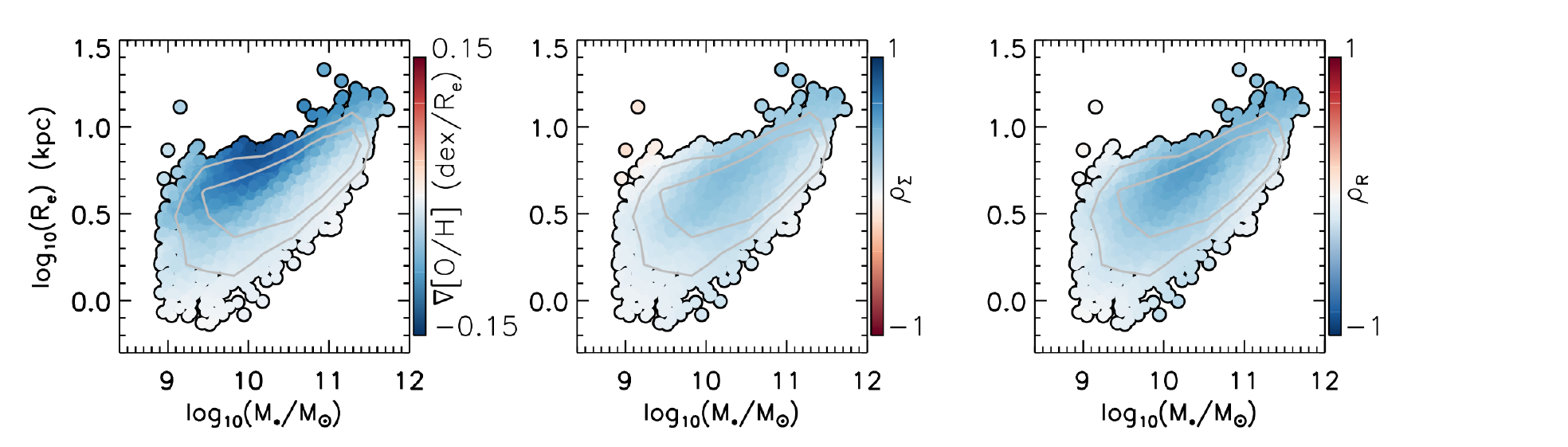}
	\caption{\rev{Effective radius plotted against galaxy stellar mass, with data points coloured by the radial gas metallicity gradients (left) and by the Spearman correlation coefficients between $\Sigma_*$ and metallicity (middle) and between radius and metallicity (right); the gas metallicity here is determined from the P16 calibrator. We apply LOESS smoothing in all panels. The middle panel uses a reversed colour scale; thus, blue regions correspond to negative gradients, positive $\Sigma_*$-metallicity correlations and negative radius-metallicity correlations. The contours enclose $\sim$50\%\ and $\sim$80\%\ of galaxies.}}
	\label{rhoradgrad_p16}
	\end{center}
\end{figure*}

\rev{Similarly to the M13 calibrator, the P16 R2 calibrator was determined from observational data; however, it employs the $\mathrm{[OII]}\lambda3737, 3729$ doublet in addition to the lines within the O3N2 indicator, making the R2 calibrator more resistant to biases from N/O variations. The R2 calibrator employs the full [OIII] and [NII] doublets, so we assume a 1/3 ratio between the dominant and subdominant lines of these doublets. We employ the same spaxel/galaxy sample presented in the main paper text. }

\begin{figure*}
\begin{center}
	\includegraphics[trim = 0.5cm 10.5cm 0cm 13cm,scale=1,clip]{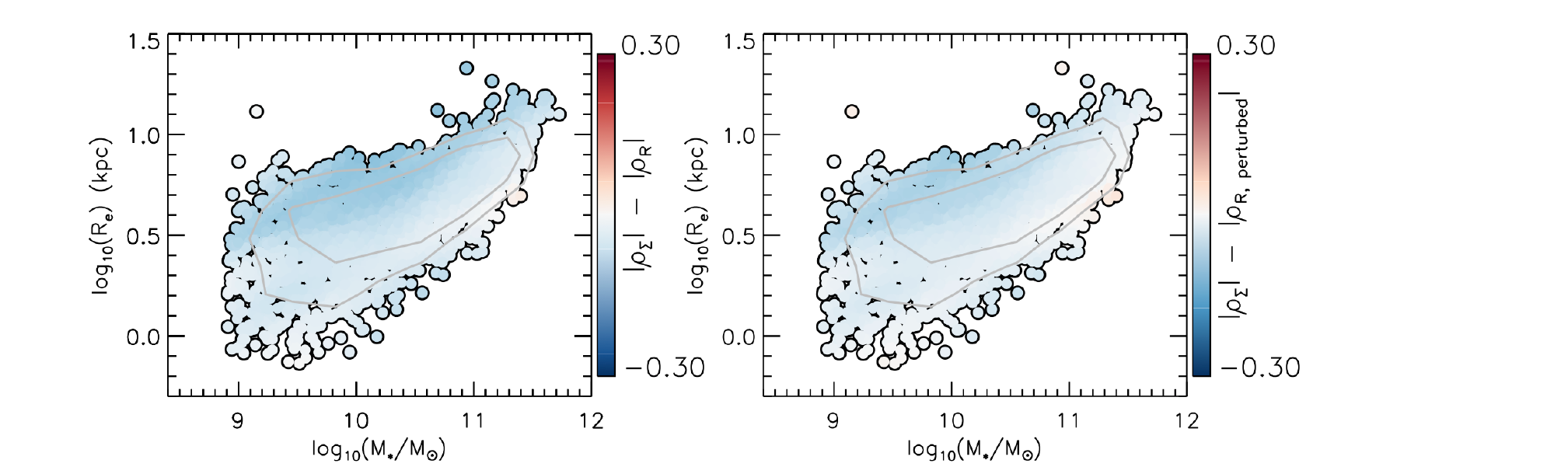}
	\caption{\rev{Difference in magnitudes of the Spearman correlation coefficient between gas metallicity and stellar density, $|\rho_{\Sigma} |$, and the coefficient between gas metallicity and galactocentric radius before ($|\rho_{R}|$; left panel) and after ($|\rho_{R, perturbed}|$; right panel) perturbing spaxel radii. Each datapoint represents one galaxy, and LOESS smoothing is applied; blue regions indicate where metallicity is more strongly correlated with radius, while red regions indicate a stronger correlation with density. The contours enclose $\sim$50\%\ and $\sim$80\%\ of galaxies.}}
	\label{corrdif_p16}
	\end{center}
\end{figure*}

\begin{figure*}
\begin{center}
	\includegraphics[trim = 1cm 11cm 0cm 13cm,scale=1,clip]{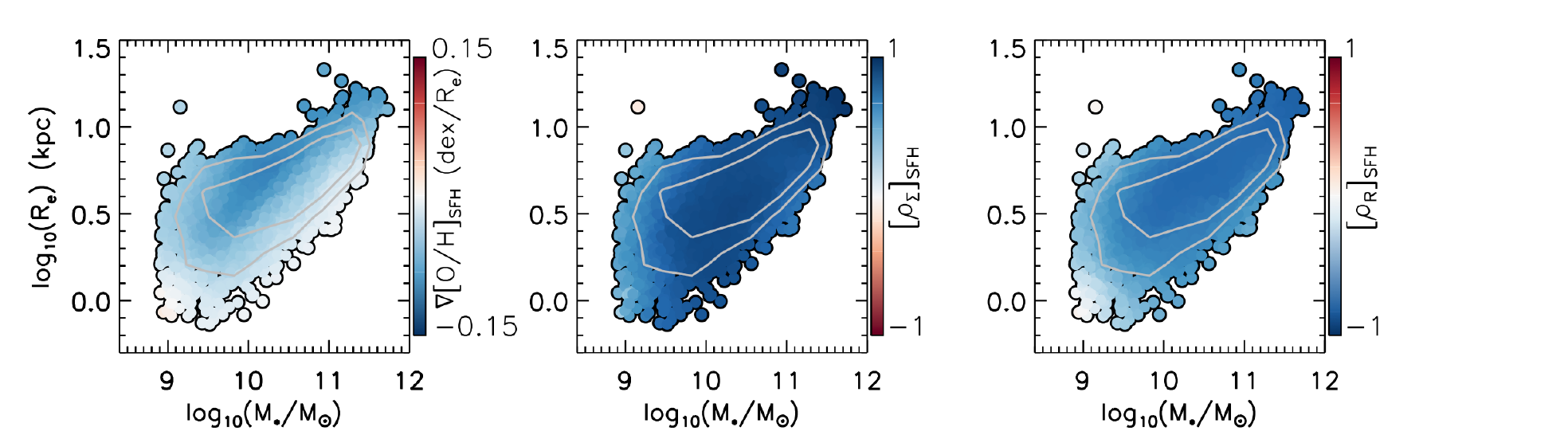}
	\caption{\rev{Effective radius, plotted against galaxy stellar mass, with data points coloured by the B22 `SFH model' metallicity gradients (left) and by the Spearman correlation coefficients between $\Sigma_*$ and model metallicity ($[\rho_{\Sigma}]_{SFH}$, middle) and between radius and model metallicity (right). The model metallicities were build using the P16 R2 indicator in this case. We apply LOESS smoothing in all panels, and the middle panel uses a reversed colour scale. The contours enclose $\sim$50\%\ and $\sim$80\%\ of galaxies. We find that the SFH models do \textit{not} reproduce the trends shown in Figures 1 and 2 for observed gas metallicities: while individual parameters behave similarly across the mass-size plane, we find local metallicties to be correlated more strongly with stellar density than with radius in most cases.}}
	\label{rhoradgrad_sfh_p16}
	\end{center}
\end{figure*}

\rev{We present in \autoref{rhoradgrad_p16} gas metallicity gradients, $\rho_\Sigma$ and $\rho_R$ values across the mass-size plane. These were calculated in an identical manner to what was described in the main paper text, but with P16 metallicities instead of M13 metallicities. We then present in \autoref{corrdif_p16} a comparison between $|\rho_\Sigma |$ and $|\rho_R|$ and between $|\rho_\Sigma |$ and $|\rho_{R, perturbed}|$, with all values calculated in the same way as before. Finally, we present in \autoref{rhoradgrad_sfh_p16} the behaviour of B22 `SFH models' as derived from the P16 calibrator.}

\rev{The correlations are found to be a little weaker here than was found from the M13 calibrator; this is unsurprising, as B22 previously noted a larger scatter in gas metallicity relations when the P16 R2 calibrator was applied. Nonetheless, we obtain the same broad results that were presented and discussed in the main paper text: $\rho_\Sigma$ and $\rho_R$ trend similarly to the gradients across the mass-size plane, as is to be expected, with $|\rho_R|$ greater than $|\rho_\Sigma |$ for less massive extended galaxies. We further find that the SFH models remain unable to replicate this behaviour, predicting stronger metallicity correlations with $\Sigma_*$ than with radius. Thus, we find that the main results from this paper are \textit{not} simply due to the use of the M13 calibrator.}

\section{Relation between gas metallicity gradients and correlation coefficients}\label{appendix_rhoradgrad}

\begin{figure*}
\begin{center}
	\includegraphics[trim = 1cm 11cm 0cm 11cm,scale=0.9,clip]{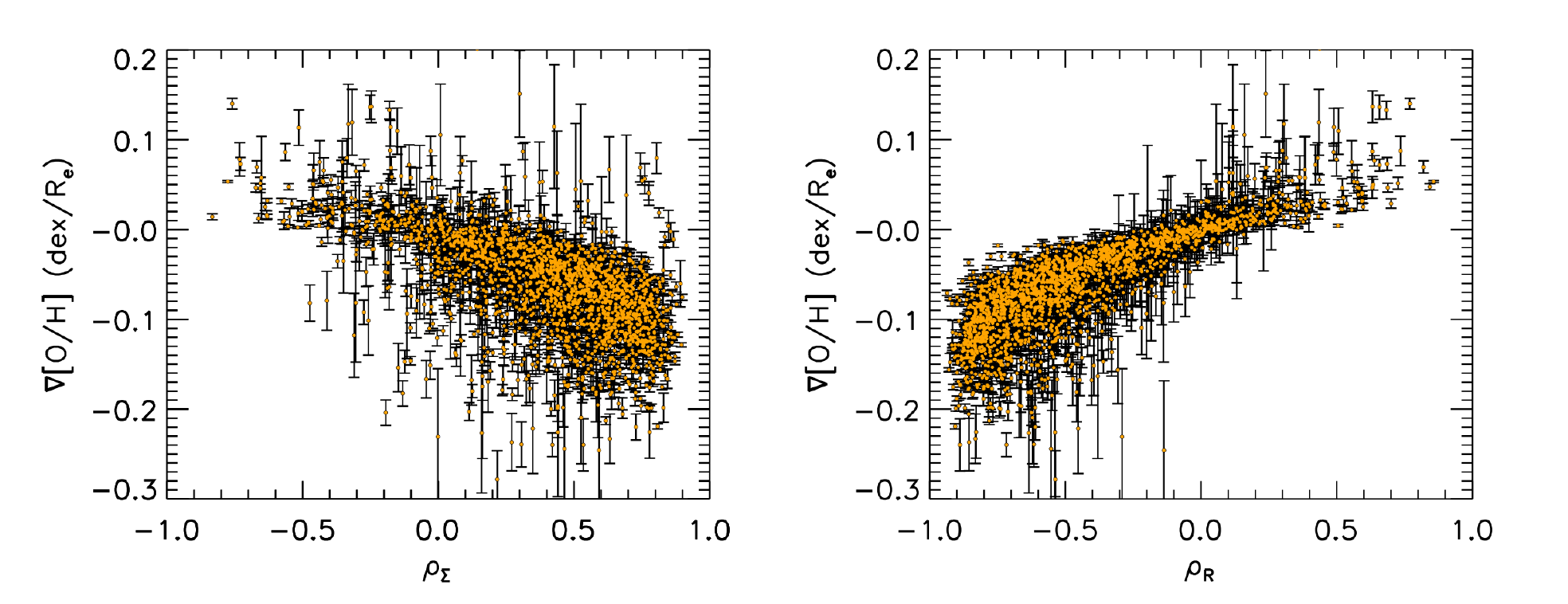}
	\caption{\rev{Radial gas metallicity gradients plotted against $\rho_\Sigma$ (left) and $\rho_R$(right), with gradients obtained using the M13 calibrator. A small number of anomalous gradient values are not shown for presentational purposes.}}
	\label{radrho_coeffs}
	\end{center}
\end{figure*}

\rev{We now explore in more depth how the gas metallicity gradients relate to $\rho_R$ and $\rho_\Sigma$, focussing specifically on the M13 calibrator. In \autoref{radrho_coeffs}, we plot both $\rho_R$ and $\rho_\Sigma$ against the radial metallicity gradient; as was already apparent, steeper gradients are associated more positive $\rho_\Sigma$ values and more negative $\rho_R$ values.}

\begin{figure*}
\begin{center}
	\includegraphics[trim = 1cm 11cm 0cm 7cm,scale=0.8,clip]{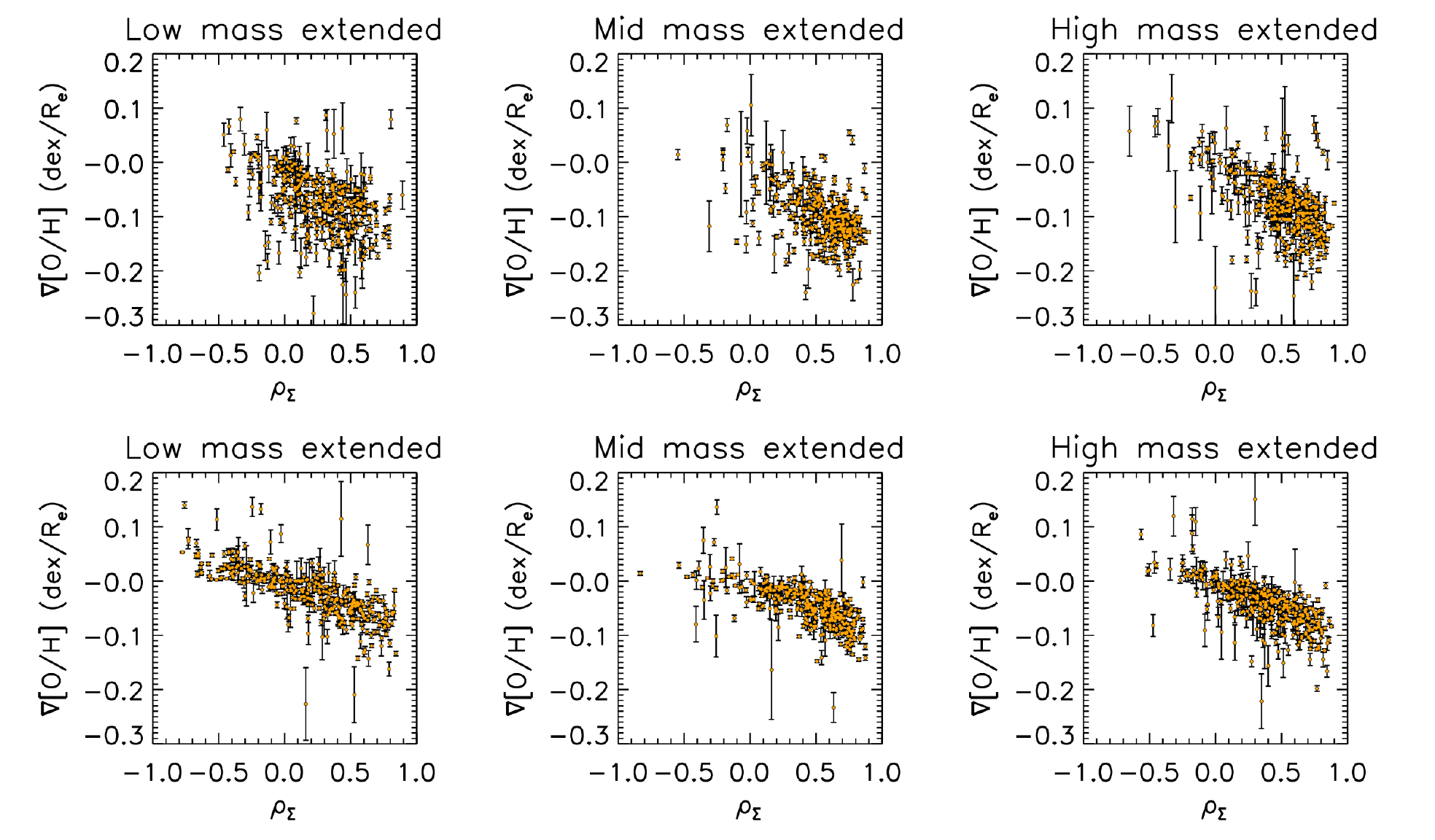}
	\caption{\rev{Radial gas metallicity gradients plotted against $\rho_\Sigma$ for the six mass-size bins first described in Section 3.2.}}
	\label{rho_coeffs_msbin}
	\end{center}
\end{figure*}

\begin{figure*}
\begin{center}
	\includegraphics[trim = 1cm 11.cm 0cm 7cm,scale=0.8,clip]{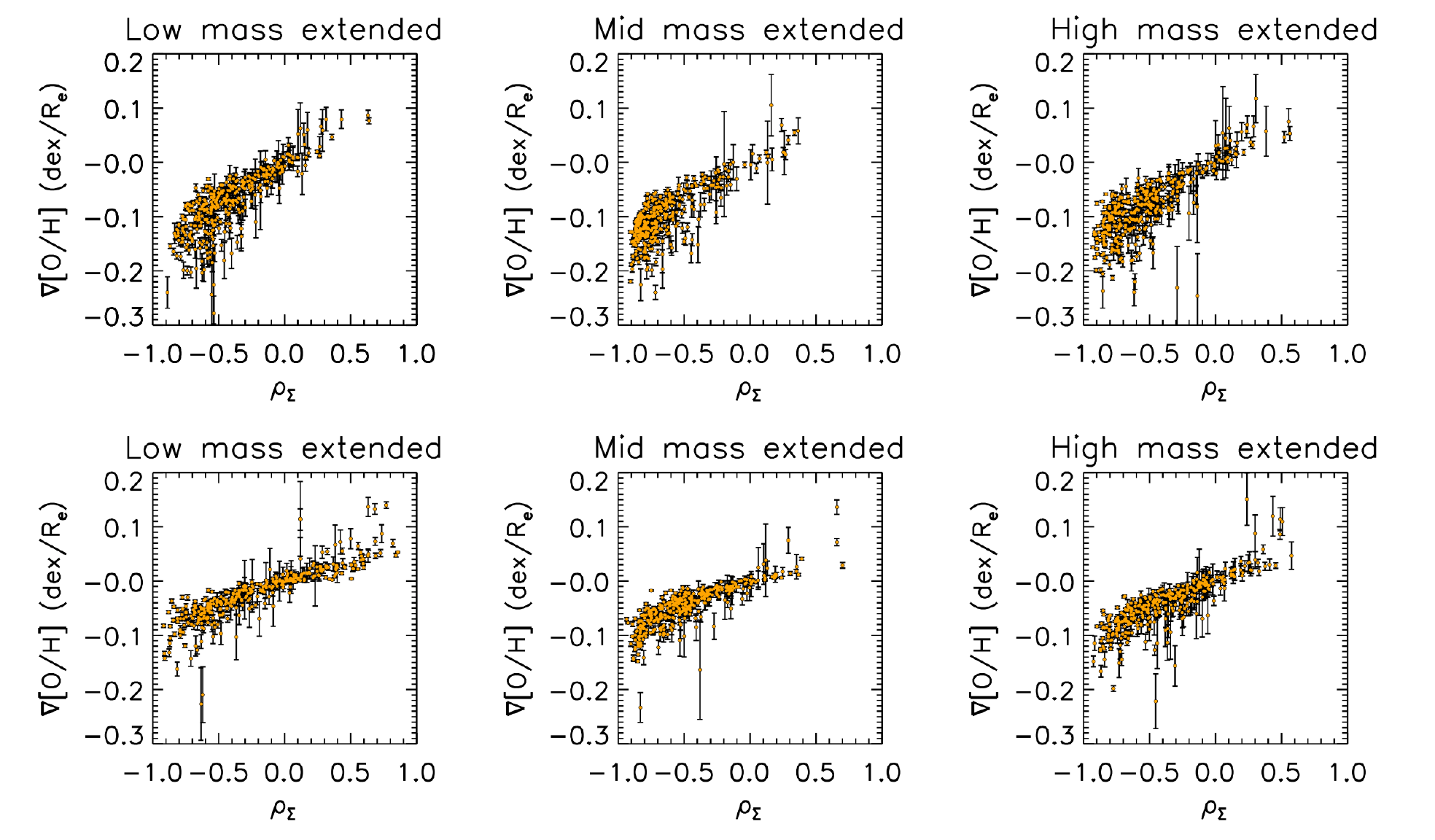}
	\caption{\rev{Radial gas metallicity gradients plotted against $\rho_R$ for the six mass-size bins first described in Section 3.2.}}
	\label{rad_coeffs_msbin}
	\end{center}
\end{figure*}

\rev{Next, we consider how metallicity gradients vary with $\rho_\Sigma$ and $\rho_R$ at different locations in the mass-size parameter space. For this, we employ the six mass-size subsamples described in Section 3.2. We plot in \autoref{rho_coeffs_msbin} the gas metallicity gradients against $\rho_\Sigma$ for each of the subsamples, while in \autoref{rad_coeffs_msbin} we compare the metallicity gradients and $\rho_R$ in the same manner.} 

\rev{For both the sample as a whole and for the mass-size subsamples, it is apparent that the gradient correlates tightly with $\rho_\Sigma$ and especially with $\rho_R$. This is by construction: $\Sigma_*$ declines with radius, meaning that steeper gradients can be immediately expected to yield tighter metallicity correlations with both $\Sigma_*$ and radius.}

\end{appendix}
\label{lastpage}
\end{document}